\documentclass[a4paper,11pt]{article}
\usepackage{pos}

\newcommand{\be}{\begin{equation}}
\newcommand{\ee}{\end{equation}}
\newcommand{\bea}{\begin{eqnarray}}
\newcommand{\eea}{\end{eqnarray}}

\newcommand{\la}{\langle}
\newcommand{\ra}{\rangle}

\title{A pion decay constant in the multi-flavor Schwinger model}

\author*[a]{Jaime Fabi\'{a}n Nieto Castellanos}
\author[b]{Ivan Hip}
\author[a]{Wolfgang Bietenholz}

\affiliation[a]{Instituto de Ciencias Nucleares \\
  Universidad Nacional Aut\'{o}noma de M\'{e}xico \\
  A.P.\ 70-543, C.P.\ 04510 Ciudad de M\'{e}xico, Mexico}

\affiliation[b]{University of Zagreb Faculty of Geotechnical Engineering\\
Hallerova aleja 7, 42000 Vara\v{z}din, Croatia}

\emailAdd{jafanica@ciencias.unam.mx}
\emailAdd{ivan.hip@gfv.unizg.hr}
\emailAdd{wolbi@nucleares.unam.mx}

\abstract{The pion decay constant $F_{\pi}$ plays an important role
  in QCD and in Chiral Perturbation Theory. It is hardly known,
  however, that a corresponding constant exists in the Schwinger
  model with $N_{\rm f} \geq 2$ degenerate fermion flavors. In this case,
  the ``pion'' does not decay and $F_{\pi}$is dimensionless. Still,
  $F_{\pi}$ can be defined by 2d analogies to the Gell-Mann--Oakes--Renner
  relation, the Witten--Veneziano formula and the residual ``pion'' mass
  in the $\delta$-regime. With suitable assumptions, and by inserting
  simulation data, these QCD-inspired relations are all compatible
  with $F_{\pi} \simeq 1/\sqrt{2\pi}$ at zero fermion mass, as we
  observe for $N_{\rm f} = 2, \dots , 6$. We conclude that this is a
  meaningful constant in the multi-flavor Schwinger model.}

\FullConference{The 40th International Symposium on Lattice Field Theory (Lattice 2023)\\
July 31st - August 4th, 2023\\
Fermi National Accelerator Laboratory\\}

\begin{document}
\maketitle

\section{The multi-flavor Schwinger model}
The Schwinger model \cite{Schwinger}, or 2d QED, shares important
qualitative features with QCD, in particular confinement \cite{Coleman},
chiral symmetry (breaking) and a topological structure of the gauge
configurations.
We are going to apply several aspects of the analogy between 2d QED and
4d QCD to define a ``pion decay constant'' $F_{\pi}$ in the Schwinger
model with $N_{\rm f} \geq 2$ degenerate fermion flavors.

Most analytic treatments are based on bosonization, which ---
for fermion mass $m=0$ --- leads to a boson with mass
$M_{\eta} = g \sqrt{N_{\rm f} / \pi}$ (where $g$ is the gauge coupling),
plus $N_{\rm f}-1$ massless ``pions'', see {\it e.g.}\ Refs.\
\cite{Belvedere}. Other works, however, such as Ref.\ \cite{GatSei},
assume $N_{\rm f}^{2}-1$ ``pions'', which matches the number of
Nambu-Goldstone bosons in higher dimensions, when the chiral
symmetry breaks spontaneously.

In this framework, we are interested in a quantity, which can be
defined as a ``pion decay constant'' $F_{\pi}$, by invoking three
different analogies to 4d QCD (although the ``pion'' in the Schwinger
model does not decay). To the best of our knowledge, the only work
which studied this constant before was carried out for $N_{\rm f}=2$
with a light-cone formulation \cite{Harada}. Referring to the
divergence of the axial current $J_{\mu}^{5}$,\,
$\la 0 | \partial^{\mu} J_{\mu}^{5}(0) | \pi (p) \ra$,
that study obtained
\be
F_{\pi} (m) = 0.394518(14) + 0.040(1) \, m/g \ .
\ee
Our results, to be summarized in the continuation, were presented in
detail in Ref.\ \cite{pap}, see also Refs.\ \cite{Jaime}.

\section{The 2d Gell-Mann--Oakes--Renner relation}
In QCD, the Gell-Mann--Oakes--Renner relation is well-known \cite{GMOR},
\be  \label{GMOR}
F_{\pi}^{2}(m) = \frac{2m \Sigma}{M_{\pi}^{2}} \ ,
\ee
where $\Sigma$ is the chiral condensate. If we assume the same
relation to hold in the multi-flavor Schwinger model, and combine it
with the relation
$\Sigma = M_{\pi}^{2} / 4 \pi m$
\cite{HosoRod}, we immediately arrive at
\be  \label{Fpitheo}
F_{\pi} = \frac{1}{\sqrt{2\pi}} \simeq 0.3989 \ .
\ee
without any mass-dependence.

\begin{figure}[h!]
\vspace*{-5mm}
\begin{center}
\includegraphics[angle=0,width=.5\linewidth]{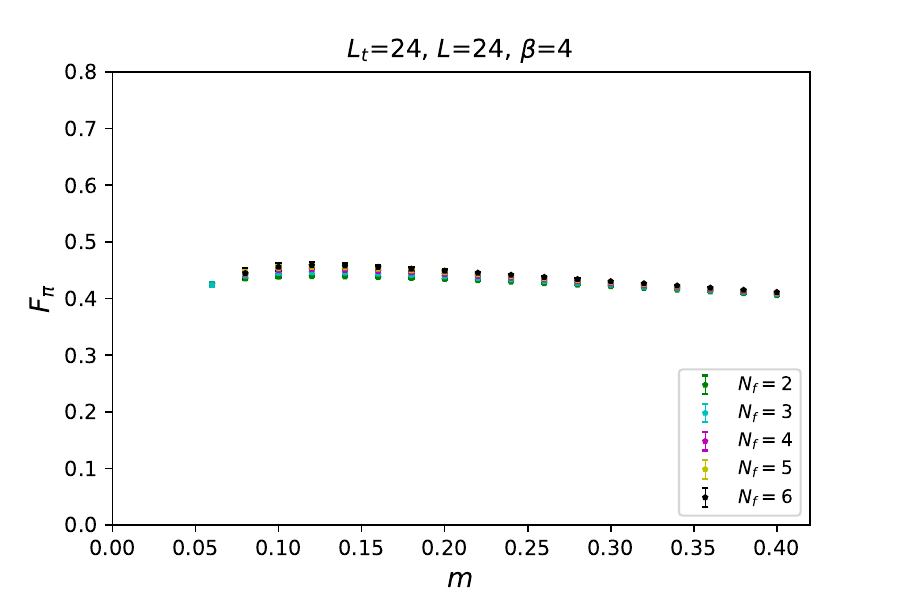}
\hspace*{-3mm}
\includegraphics[angle=0,width=.5\linewidth]{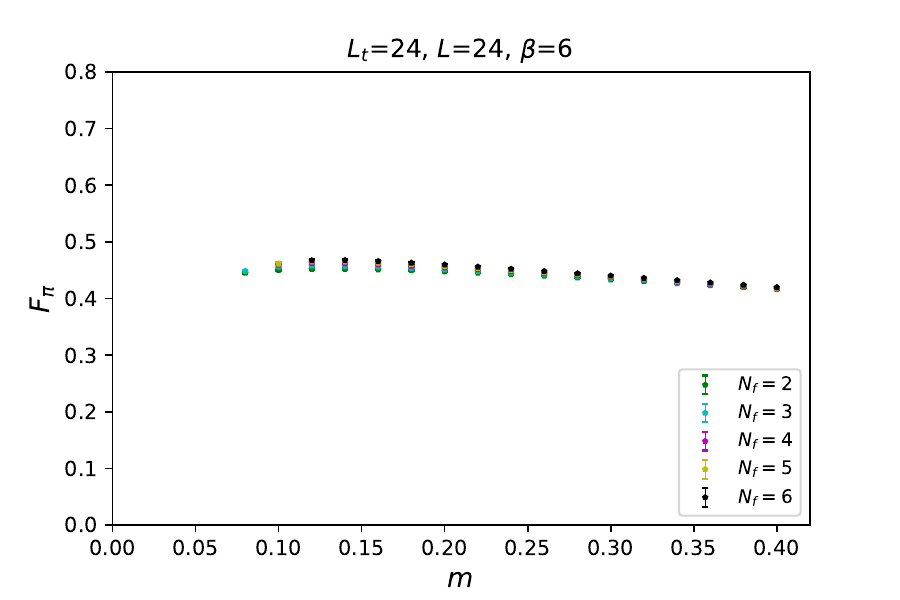}
\end{center}
\vspace*{-6mm}
\caption{Results for $F_{\pi}(m)$ based on the Gell-Mann--Oakes--Renner
  relation (\ref{GMOR}). We are using overlap-hypercube fermions,
  so we can insert for $m$ the bare fermion mass,
  while $M_{\pi}$ and $\Sigma$ are measured with
  quenched re-weighted configurations
  (for $\Sigma$ we use the Dirac spectrum).
  The results at $\beta=4$ (left)
  and  $\beta=6$ (right) are very similar, which shows that lattice
  artifacts are mild. Finite-size effects could be an issue, for this
  reason we exclude tiny fermion masses $m$. Still, we see that
  $F_{\pi} (m\to 0) \approx 0.4$.}
\vspace*{-1mm}
\label{GMORfig}
\end{figure}

Alternatively, we can numerically measure the terms in the
Gell-Mann--Oakes--Renner relations. We do so on $24 \times 24$
lattices, by using overlap-hypercube fermions \cite{WB},
which are treated by re-weighting quenched configurations,
for $N_{\rm f} = 2, \dots , 6$.
The results at $\beta =4$ and $6$ are very similar, as Figure
\ref{GMORfig} shows. We truncate the results at $m \leq 0.05$,
to avoid strong finite-size effects, but we see that the chiral
extrapolation $m \to 0$ is in all cases compatible
with $F_{\pi}(0) \approx 0.4$.

\section{Witten--Veneziano formula in the Schwinger model}
According to Seiler and Stamatescu \cite{SeiSta}, the famous
Witten--Veneziano formula \cite{WitVen}
\be
M_{\eta}^{2} = \frac{2N_{\rm f}}{F_{\eta}^{2}} \chi_{\rm t}^{\rm q}
\ee
is on particularly solid grounds in the multi-flavor Schwinger model
at $m=0$.
$M_{\eta}$ is given in Section 1, and
the quenched topological susceptibility (in infinite volume) amounts
to $\chi_{\rm t}^{\rm q} = g^{2}/4 \pi^{2}$ \cite{SeiSta}, which is
confirmed analytically and numerically by the continuum limits
of different lattice formulations \cite{chitq,pap}. This leads
to the $\eta$-decay constant $F_{\eta} = 1/\sqrt{2\pi}$.

In large-$N_{\rm c}$ QCD, $F_{\pi}$ and $F_{\eta}$ coincide asymptotically.
If we drive the analogy further and assume the same equivalence in the
multi-flavor Schwinger model, we obtain
\be  \label{Fpitheo2}
F_{\pi}(m=0) = \frac{1}{\sqrt{2\pi}} \ ,
\ee
in agreement with eq.\ (\ref{Fpitheo}), though here with the
limitation to the chiral limit.

\section{The 2d $\delta$-regime}
One formulation of Chiral Perturbation Theory refers to the
$\delta$-regime, {\it i.e.}\ to an anisotropic space-time volume
with $L_{t} \gg L \approx M_{\pi}^{-1}$ \cite{Leutwyler}. Thus the system
is quasi-1-dimensional and it can be approximated by a quantum mechanical
rotor. In the chiral limit of zero quark masses, there is still a residual
pion mass $M_{\pi}^{\rm R}$, as a finite-size effect,
\be  \label{Mpires}
M_{\pi}^{\rm R} = \frac{N_{\pi}}{2 \Theta_{\rm eff}} \ ,
\ee
where $N_{\pi}$ is the number of pions (3 in QCD), and
$\Theta_{\rm eff}$ is an effective moment of inertia. To leading
order (LO), Leutwyler computed $\Theta_{\rm eff} = F_{\pi}^{2} L^{3}$
\cite{Leutwyler}. For numerical studies with lattice QCD, we refer
to Ref.\ \cite{Bietenholz2010}. Hasenfratz and Niedermayer extended
this calculation to next-to-leading order (NLO) of an O($N$) model
in $d>2$ \cite{Hasenfratz93},
\be
\Theta_{\rm eff} = F_{\pi}^{2} L^{d-1} \Big[ 1 +
  \frac{N_{\pi}-1}{2 \pi F_{\pi}^{2} L^{d-2}} \Big( \frac{d-1}{d-2} + \dots
  \Big) + \dots \Big]  \ .
\ee
They assumed spontaneous symmetry breaking ${\rm O}(N) \to {\rm O}(N-1)$,
and therefore $N_{\pi} = N-1$.

In $d=2$ this does not happen, and the NLO correction would diverge,
so we can only conjecture that the LO remains applicable. If a numerical
study in the multi-flavor Schwinger model confirms the behavior
$M_{\pi}^{\rm R} \propto 1/L$, then eq.\ (\ref{Mpires}) provides another
result for $F_{\pi}$.

We performed such simulations in two settings, with $10^{4}$
configurations for each parameter set:

\begin{itemize}

\item Dynamical Wilson fermions, using the HMC algorithm,
  with $N_{\rm f}=2$, $L_{t}=64$,
  $L= 6, \dots , 12$ and $\beta \equiv 1/g^{2}= 3,\, 4$ and $5$.
  
\item Overlap-hypercube fermions, with quenched configurations
and re-weighting for $N_{\rm f}= 2, \dots , 6$, with $L_{t}=32$
and $L=4, \dots , 12$, at $\beta=4$.
  
\end{itemize} 

\begin{figure}[h!]
\vspace*{-5mm}
\begin{center}
\includegraphics[angle=0,width=.5\linewidth]{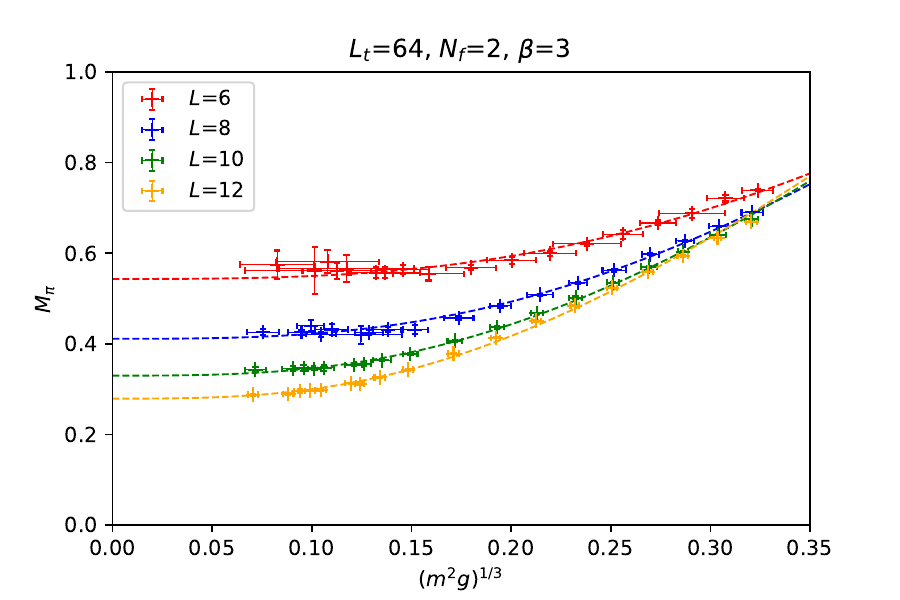}
\hspace*{-3mm}
\includegraphics[angle=0,width=.5\linewidth]{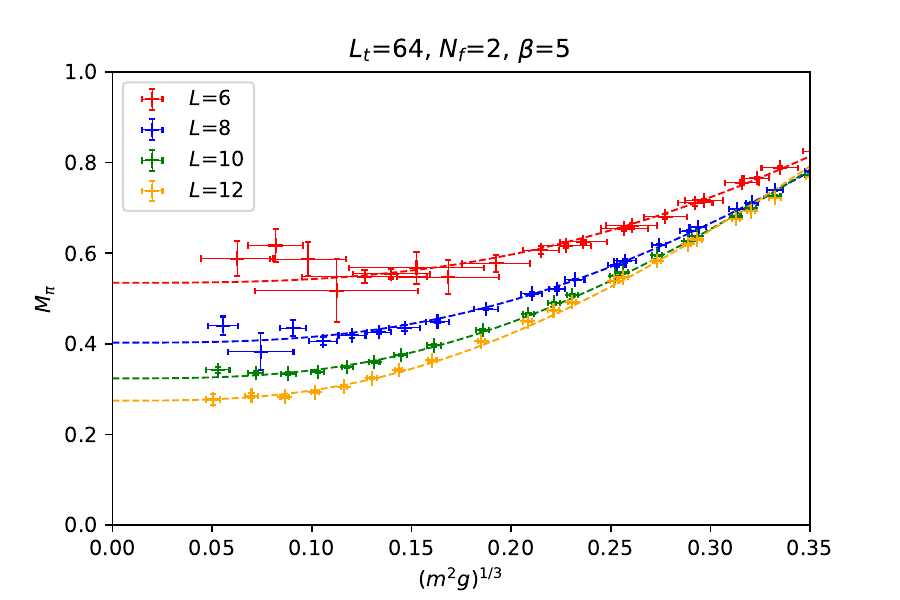}
\includegraphics[angle=0,width=.5\linewidth]{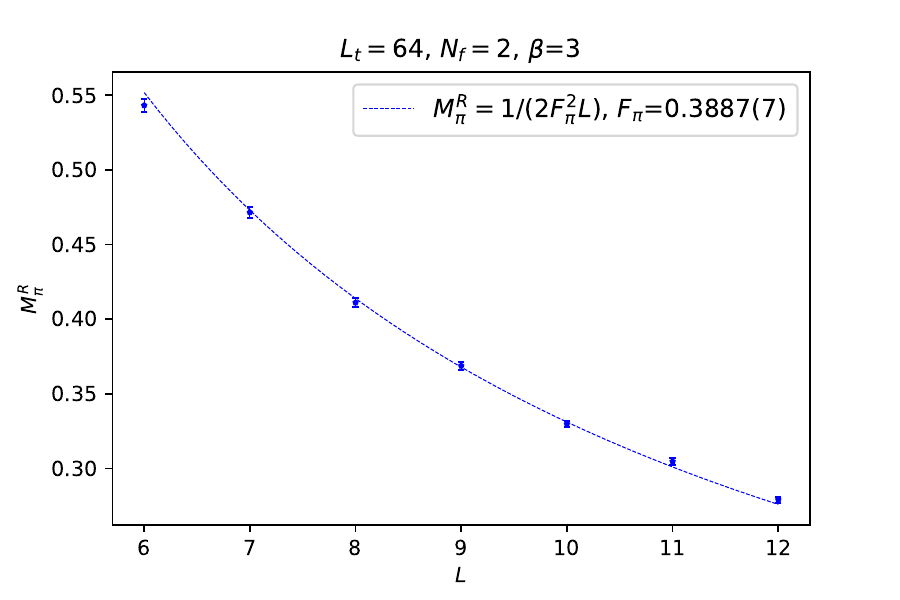}
\hspace*{-3mm}
\includegraphics[angle=0,width=.5\linewidth]{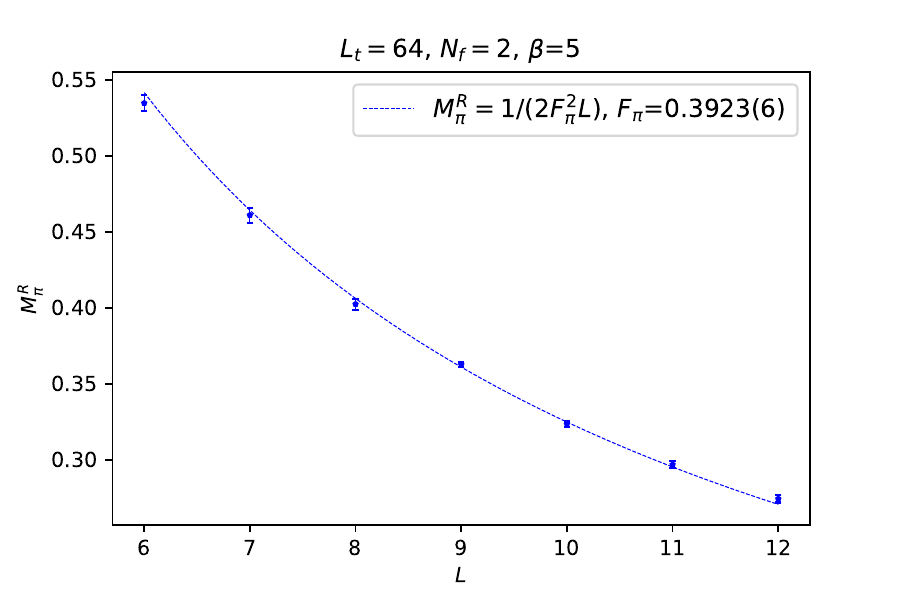}
\end{center}
\vspace*{-6mm}
\caption{Results for $M_{\pi}$ from simulations in the $\delta$-regime,
with Wilson fermions and $N_{\rm f}=2$, at $\beta=3$ and $5$. Above
we show $M_{\pi}$ as a function of $(m^{2}g)^{1/3}$, with the extrapolations
to $M_{\pi}^{\rm R} = M_{\pi}(m=0)$. The plots below confirm the behavior
$M_{\pi}^{\rm R} \propto 1/L$, and the resulting values for $F_{\pi} (m=0)$
are again close to $1/\sqrt{2\pi}$.}
\vspace*{-1mm}
\label{deltaWilson}
\end{figure}

Figure \ref{deltaWilson} summarizes our results for $N_{\rm f}=2$
with Wilson fermions.
The plots above show $M_{\pi}$ as a function of the parameter
$(m^{2}g)^{1/3}$ (in lattice units), at $\beta \equiv 1/g^{2} =3$
and $5$. The (degenerate) fermion mass $m$ is measured
based on the PCAC relation, as it was done previously
{\it e.g.}\ in Ref.\ \cite{GattHipLang}. As expected for Wilson
fermions, the uncertainty becomes significant at small
values of $m$, but there are smooth extrapolations to
$M_{\pi}^{\rm R} = M_{\pi}(m=0)$.

The plots below show how this residual pion mass $M_{\pi}^{\rm R} $
depends on $L$.
The relation $M_{\pi}^{\rm R} \propto 1/L$ is well confirmed, which
allows us to extract a value for $F_{\pi}$ according to eq.\
(\ref{Mpires}). Here we insert $N_{\pi} =1$, which matches the
formula used in bosonization studies, as we mentioned in Section 1.
This leads to results for $F_{\pi} (m=0)$ which are again close to
eq.\ (\ref{Fpitheo2}). Our results obtained in this manner for
$\beta =3,\, 4$ and $5$ are given in Table \ref{tabFpidelta}.
They show that lattice artifacts are small, and they provide a
picture of the chiral limit, which is consistent with Sections 2 and 3.

\begin{table}[h!]
 \begin{center}
  \begin{tabular}{|c||c|c|c|}
        \hline   
	$\beta = 1/g^{2} $ & 3 & 4 & 5 \\
        \hline
        $F_\pi$  & 0.3887(7) & 0.3877(11) & 0.3923(6) \\
	\hline
  \end{tabular}
 \end{center}
 \vspace*{-2mm}
 \caption{Results for $F_{\pi}$, obtained from simulations of
 two flavors of Wilson fermions in the $\delta$-regime,
 by fits to eq.\ (\ref{Mpires}), with $N_{\pi}=1$,
 at three values of $\beta$.}
 \label{tabFpidelta}
 \vspace*{-2mm}
\end{table}

We now extend our study in the $\delta$-regime
to $N_{\rm f} = 2, \dots , 6$ degenerate fermion flavors.
Here we use quenched configurations, generated at $\beta=4$,
which are re-weighted with the fermion determinant which
corresponds to overlap-hypercube fermions.

\begin{figure}[h!]
\begin{center}
\includegraphics[angle=0,width=.5\linewidth]{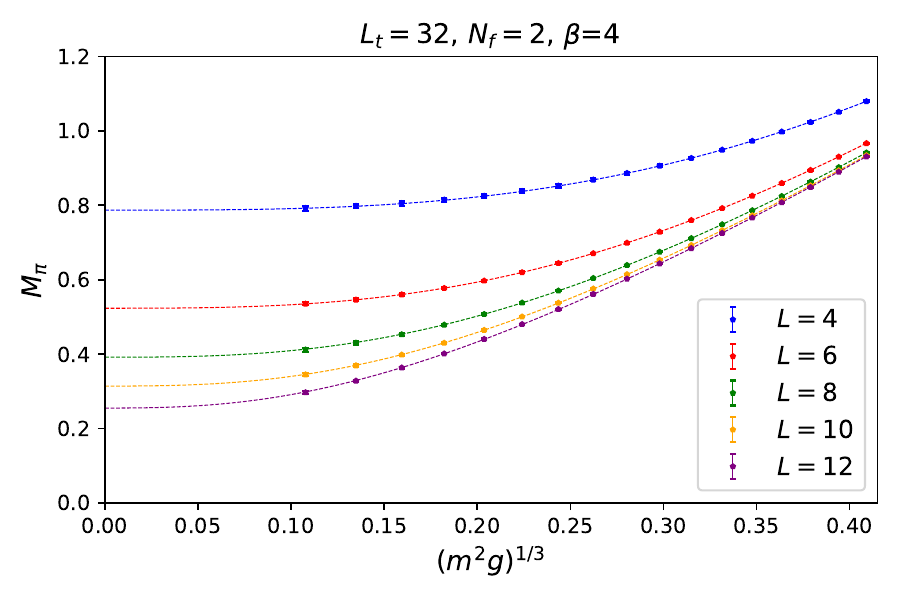}
\hspace*{-3mm}
\includegraphics[angle=0,width=.5\linewidth]{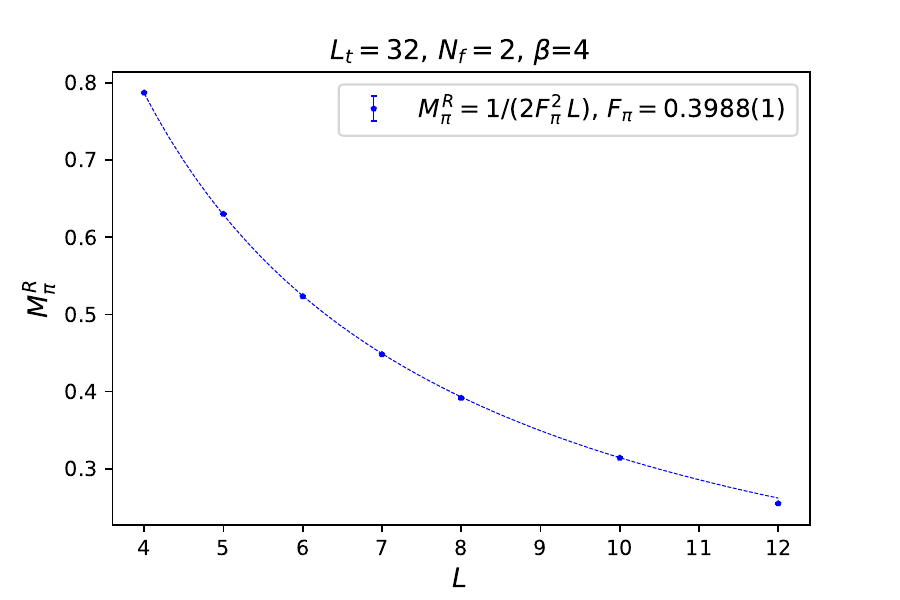}
\end{center}
\vspace*{-6mm}
\caption{Results for $M_{\pi}$ from simulations in the $\delta$-regime,
with overlap-hypercube fermions and $N_{\rm f}=2$, at $\beta=4$. On the left
we show $M_{\pi}$ as a function of $(m^{2}g)^{1/3}$, with the extrapolations
to $M_{\pi}^{\rm R} = M_{\pi}(m=0)$. The plot on the right confirms the behavior
$M_{\pi}^{\rm R} \propto 1/L$, and the resulting value for $F_{\pi} (m=0)$
is once more close to $1/\sqrt{2\pi}$.}
\vspace*{-1mm}
\label{deltaoverlapHF1}
\end{figure}

Figure \ref{deltaoverlapHF1} shows, for $N_{\rm f}=2$, the ``pion'' mass
$M_{\pi}((m^{2}g)^{1/3})$ (again we can use the bare mass $m$ thanks
to the chiral symmetry of Ginsparg-Wilson fermions)
and $M_{\pi}^{\rm R}(L)$, as in Figure \ref{deltaWilson}.

\begin{figure}[h!]
\begin{center}
\includegraphics[angle=0,width=.65\linewidth]{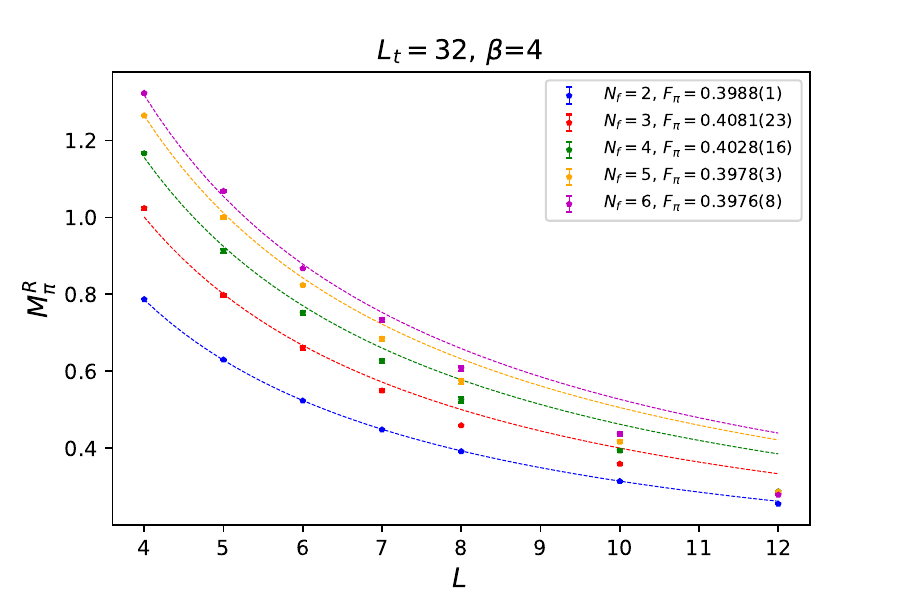}
\end{center}
\vspace*{-6mm}
\caption{Results for $M_{\pi}$ from simulations in the $\delta$-regime,
with overlap-hypercube fermions and $N_{\rm f}=2, \dots ,6$, at $\beta=4$.
We see that increasing $N_{\rm f}$ reduces the range where the relation
$M_{\pi}^{\rm R} \propto 1/L$ is well approximated. Fits within this
range, along with the application of the effective formula (\ref{magic}),
lead to  values for $F_{\pi} (m=0)$, which are again close to
$1/\sqrt{2\pi}$.}
\vspace*{-1mm}
\label{deltaoverlapHF2}
\end{figure}

Finally, Figure \ref{deltaoverlapHF2} presents the corresponding
results if the re-weighting is performed for $N_{\rm f} = 2,\dots ,6$.
We see that increasing $N_{\rm f}$ reduces the range in $L$
where the proportionality relation $M_{\pi}^{\rm R} \propto 1/L$
is well approximated. We perform fits within this range
for each $N_{\rm f}$.

Now the question is what value for $N_{\pi}$ should be inserted in
eq.\ (\ref{Mpires}). The bosonization formula $N_{\pi} = N_{\rm f}-1$
fails for $N_{\rm f} >2$. However, we obtain values for $F_{\pi} (m=0)$
which are well consistent with the previous considerations, if we
apply the effective formula
\be  \label{magic}
N_{\pi} = \frac{N_{\rm f}-1}{N_{\rm f}} \ ,
\ee
although these values for $N_{\pi}$ are non-integers for $N_{\rm f} >2$.
The results for $F_{\pi} (m=0)$ are displayed inside Figure 
\ref{deltaoverlapHF2}.

\section{Conclusions}
We have attracted attention to a constant, which plays an
important role in the multi-flavor Schwinger model
(with $N_{\rm f} \geq 2$), but
which has been ignored in the literature, with the
exception of the light-cone study in Ref.\ \cite{Harada}.

In several respects, this constant is analogous to the
pion decay constant in QCD and Chiral Perturbation Theory,
hence we denote it as $F_{\pi}$. It is, however, dimensionless
in $d=2$.

We presented three ways to introduce this constant, which
all involve some analogy between the Schwinger model and QCD.
They refer to the Gell-Mann--Oakes--Renner relation, to the
Witten--Veneziano formula and to the residual pion mass in
the $\delta$-regime of a small spatial box, but a large extent
in Euclidean time. For details, we refer to Ref.\ \cite{pap}.

All three considerations lead to results close to
$F_{\pi} (m=0) = 1/\sqrt{2 \pi} \simeq 0.3989$, which is
also close to the value obtained in Ref.\ \cite{Harada}.
We conclude that this constant is physically significant.
Further aspects of its meaning remain to be explored.

\noindent
\ \\
{\bf Acknowledgments:}
We thank Stephan D\"{u}rr, Christian Hoelbling
and Satoshi Iso for helpful comments.
The code was developed at the cluster Isabella of the Zagreb University
Computing Centre (SRCE), and the production runs were performed at
the cluster of the Instituto de Ciencias Nucleares, UNAM.
This work was supported by the Faculty of Geotechnical Engineering
of Zagreb University through the project ``Change of the Eigenvalue
Distribution at the Temperature Transition'' (2186-73-13-19-11),
by UNAM-DGAPA through PAPIIT projects IG100219 and IG100322, and
by the Consejo Nacional de Humanidades, Ciencia y Tecnolog\'{\i}a
(CONAHCYT).

\end{document}